\documentclass[journal]{IEEEtran}

\usepackage{amsmath,amssymb,amsfonts}
\usepackage{stmaryrd}
\usepackage{booktabs}
\usepackage{algorithm}
\usepackage{algorithmic}
\usepackage{graphicx}
\usepackage{subfigure} 
\usepackage{float}
\usepackage{color}
\usepackage{comment}
\usepackage{caption}
\usepackage{soul,xcolor}
\usepackage{cite}
\usepackage{multirow}
\usepackage{hyperref}[colorlinks,
            linkcolor=red,
            anchorcolor=blue,
            citecolor=green
            ]
            
\usepackage[symbol]{footmisc}

\makeatletter

\newcommand{\Rmnum}[1]{\expandafter\@slowromancap\romannumeral #1@}

\makeatother

\ifCLASSINFOpdf
\else
\fi

\begin{document}

\title{Trustworthy Sensor Fusion against Inaudible Command Attacks in Advanced Driver-Assistance Systems}

\author{Jiwei~Guan,
        Lei~Pan,~\IEEEmembership{Member,~IEEE,}
        Chen~Wang,
        % ~\IEEEmembership{Member,~IEEE,}
        Shui~Yu,~\IEEEmembership{Fellow,~IEEE,}
        Longxiang~Gao,~\IEEEmembership{Senior Member,~IEEE,}
        Xi~Zheng,~\IEEEmembership{Member,~IEEE}
        % <-this % stops a space
\thanks{Jiwei~Guan and Xi~Zheng are with the School of Computing, Macquarie University, Sydney, Australia, e-mail: jiwei.guan@mq.edu.au, james.zheng@mq.edu.au. Jiwei~Guan is also with the Data61, CSIRO, Sydney, Australia, e-mail: wayne.guan@data61.csiro.au.}
\thanks{Lei~Pan is with the School of Information Technology, Deakin University, Waurn Ponds, Australia, e-mail: l.pan@deakin.edu.au.}
\thanks{Chen~Wang is with the Data61, CSIRO, Sydney, Australia.}
\thanks{Shui~Yu is with the Faculty of Engineering and Information Technology, University of Technology Sydney, Sydney, Australia.}
\thanks{Longxiang~Gao is with the Shangdong Computer Science Center, Qilu University of Technology, China, e-mail: gaolx@sdas.org.}
% <-this % stops a space
\thanks{Longxiang~Gao and Xi~Zheng are the corresponding authors.}
}

\markboth{IEEE Internet of Things Journal}%
{Guan \MakeLowercase{\textit{et al.}}: Trustworthy Sensor Fusion against Inaudible Command Attacks in Advanced Driver-Assistance Systems}

\maketitle

\begin{abstract}
There are increasing concerns about malicious attacks on autonomous vehicles. In particular, inaudible voice command attacks pose a significant threat as voice commands become available in autonomous driving systems. How to empirically defend against these inaudible attacks remains an open question. Previous research investigates utilizing deep learning-based multimodal fusion for defense, without considering the model uncertainty in trustworthiness. As deep learning has been applied to increasingly sensitive tasks, uncertainty measurement is crucial in helping improve model robustness, especially in mission-critical scenarios. In this paper, we propose the \underline{M}ultimodal \underline{F}usion \underline{F}ramework (MFF) as an intelligent security system to defend against inaudible voice command attacks. MFF fuses heterogeneous audio-vision modalities using VGG family neural networks and achieves the detection accuracy of 92.25\% in the comparative fusion method empirical study. Additionally, extensive experiments on audio-vision tasks reveal the model's uncertainty. Using Expected Calibration Errors, we measure calibration errors and Monte-Carlo Dropout to estimate the predictive distribution for the proposed models. Our findings show empirically to train robust multimodal models, improve standard accuracy and provide a further step toward interpretability. Finally, we discuss the pros and cons of our approach and its applicability for Advanced Driver Assistance Systems.
\end{abstract}

% Note that keywords are not normally used for peer review papers.

\IEEEpeerreviewmaketitle

\section{Introduction}
\label{sec:introduction}

Recent work has shown that inaudible voice commands can attack smart devices if equipped with micro-electromechanical systems (MEMS) microphones \cite{roy2017backdoor,zhang2017dolphinattack,schonherr2018adversarial}. Inaudible voice commands are higher frequency sound waves that are audible to MEMS microphones but inaudible to human-being. Inaudible voice commands are deceptive signals and are too difficult to be identified in reality. The adversary sends malicious inaudible voice commands to silently controls voice assistants such as Google Assistant and Amazon Alexa for gaining access accounts and systems, which is a severe threat. In addition, voice control is utilized by Advanced Driver Assistance Systems (ADAS) for Level-5 fully autonomous driving in the foreseeable future, but voice assistants with MEMS microphones are vulnerable to inaudible voice commands~\cite{zhang2017dolphinattack}. Such inaudible commands are designed for malicious intents, resulting in significant consequences such as accidental deaths. This attack makes autonomous driving vehicles easily targeted by adversaries. This type of attack raises an important security challenge and should be addressed urgently in ADAS. The existing approaches lack inaudible samples and require complex signal-processing techniques. Since there is no clear solution against inaudible voice commands in ADAS, this paper motivates to bridge this gap with trustworthy audio and vision modalities fusion.

\begin{figure}[!ht]
\centering
\includegraphics[width=9cm,height=4.5cm]{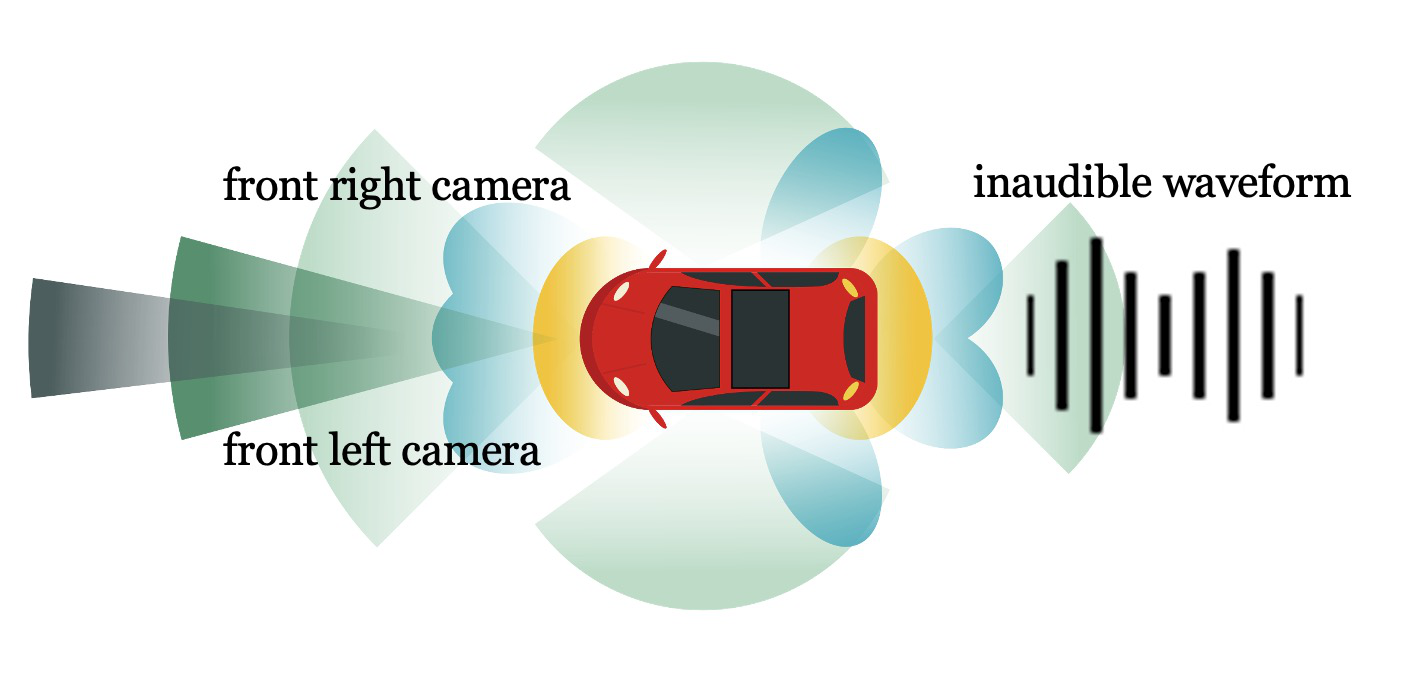}
\caption[multiple cameras]{Multiple front cameras help defend against inaudible command attacks. Inaudible commands can be designed to modulate voice commands on ultrasonic carriers. An adversary drives or stands on the road, using a high-frequency speaker to emit inaudible voice commands to attack microphones on ADAS. MFF utilizes visual evidence through cameras to defend against inaudible voice command attacks. It considers a multimodal data perspective for safe autonomous driving.}
\label{fig 1} 
\end{figure}

To the best of our knowledge, there has been little work seeking to detect inaudible attacks by cross-modal interactions. Earlier research efforts~\cite{guan2021robust} have proposed that traffic signs can be used as visual evidence to counter specific inaudible voice commands. However, the proposed defense is based on the traffic signs~\cite{stallkamp2011german} dataset, which is not strong visual evidence. Note that multiple cameras are widely used in autonomous driving perceptions, and multimodal sensor fusion is vital for a comprehensive understanding of road conditions and increasing accuracy to ensure autonomous driving~\cite{yogamani2019woodscape}. MFF takes this idea, moving a step further to detect inaudible voice command attacks. Our defense solution assumes that multi-visual semantic representations align with voice command content. The multiple visual sensors consisting of the front left and front right cameras are shown in Figure~\ref{fig 1}. The voice command will be executed if voice commands and traffic images have the same semantic content, otherwise rejected. Combining voice commands and traffic views is essential to fuse audio and vision data effectively. The proposed approach senses real/physical traffic information and monitors the driving surroundings to increase safety and security at a lower cost against inaudible voice command attacks.

Moreover, we go beyond the above work to consider the trustworthiness of multimodal fusion models. Deep learning multimodal models can be unreliable in real-world applications and standard accuracy evaluation alone is insufficient to provide trustworthy analysis. This raises deep learning trustworthy issues and how to measure model trustworthiness in mission-critical scenarios. Model uncertainty~\cite{ovadia2019uncertainty} incorporates the trustworthiness and has been studied to reveal deep learning model capabilities to provide a predictive approximation range at the test stage except for the Softmax probability. With model uncertainty awareness, users clearly understand model limitations in high-stakes Internet of Things (IoT) applications such as autonomous driving. Furthermore, developers continuously utilize such metrics to monitor deep learning models to decide whether these models can be trusted or not. Model uncertainty is generally interpreted to review the algorithmic bias output and adjust the relevant training strategy. Thus, model uncertainty is a key factor influencing the public's trust in AI. Our major contributions are listed as follows:

\begin{itemize}
\item We propose a robust audio and vision fusion defense system to effectively and accurately detect inaudible voice command attacks in a typical IoT application.

\item We build the fusion detection model based on ADAS semantic consistency between vision and audio data and present performance and ablation analysis.

\item We further provide the model uncertainty quantification metric Expected Calibration Error and uncertainty estimation Monte-Carlo Dropout for the proposed MFF.
\end{itemize}

The remainder of this paper is structured as follows: Section~\ref{sec:relatedwork} describes related work on voice attacks and ADAS attacks before describing sensor fusion and model uncertainty. Section~\ref{sec:Methodology} highlights the defense system's aims and describes how to measure model uncertainty for trustworthiness. Section~\ref{sec:Experiment} explains the data pre-processing and reports experiments settings. Section~\ref{sec:Evaluation} presents the experimental evaluation results and model uncertainty analysis. Section~\ref{sec:Discussion} discusses our proposed defense system. Finally, Section~\ref{sec:Conclusion} concludes this work and highlights interesting research directions in ADAS.

\section{Related Work} 
\label{sec:relatedwork}

\subsection{Audio Attacks}
Vaidya \textit{et al.}~\cite{vaidya2015cocaine} created audio samples acceptable to Automatic Speech Recognition (ASR) but non-meaningful for human understanding. The technique extracted acoustic information from both human and mangled voice commands. Hidden voice commands \cite{carlini2016hidden} were used to generate simple noise audio that was inaudible to humans but is effectively recognized by ASR. Furthermore, Feng \textit{et al.}~\cite{feng2017continuous} used the VAuth system to collect body-surface vibrations to match the speech signal that was received by a microphone. Scardapane \textit{et al.}~\cite{scardapane2017use} explored different Recurrent Neural Network (RNN) architectures to detect spoofing in the Automatic Speaker Verification (ASV) Spoof 2015 Challenge. As an extension of a prior study, Kreuk \textit{et al.}~\cite{kreuk2018fooling} discovered an ASV end-to-end neural network model that was vulnerable to adversarial example attacks: speech sample with noise was authenticated by the sound of speaker B, but it was generated in the sound of speaker A. Another experiment~\cite{sun2018training} explored using adversarial examples to train a more robust ASR with deep neural network-based acoustic models, achieving a 23\% word error rate reduction in the Aurora-4 database. Wang \textit{et al.}~\cite{wang2020targeted} carefully designed a Generative Adversarial Network (GAN) to generate audio adversarial examples to misclassify a speech classification network. Fundamentally, generating adversarial audio samples for ASR has been proved difficult.

Previous studies posit that an attacker needs to master knowledge about the speech recognition workflow or deep learning-based feature extraction in speech recognition. Such a white-box attack is not very practical for voice control systems. Additionally, the traditional speech recognition framework is not entirely designed by an end-to-end deep neural networks, making it difficult to study its vulnerabilities. A question arises: How do we apply inaudible voice command attacks in practice? \textit{CommanderSong}~\cite{yuan2020commandsong} injected noise voice commands into songs and passed into Kaldi~\cite{povey2011kaldi} speech recognition without being recognized by humans. Researchers conducted a dolphin attack~\cite{zhang2017dolphinattack} to hide voice commands to fool Alex and Siri systems. They replayed speech recordings through ultrasonic carriers to launch the attack. Song \textit{et al.}~\cite{song2017poster} successfully launched inaudible voice attacks through ultrasound to Micro-electromechanical systems (MEMS) microphones over a short distance. Sugawara \textit{et al.}~\cite{sugawara2020light} carried out an injection attack converting light to sound by shining a laser on the microphone's aperture. Compared with the above audio attacks, inaudible attacks are challenging to generate and implement. Inaudible attacks are a real threat because inaudible signals are stealthily emitted and hard to capture.

Adversarial examples attack deep learning-based ASR systems by adding perturbations to waveforms~\cite{qin2019imperceptible, alzantot2018did}. Carlini and Wagner~\cite{carlini2018audio} constructed targeted audio adversarial examples to attack an end-to-end ASR system~\cite{hannun2014deep, amodei2016deep}. Hu~\textit{et al.}~\cite{hu2019adversarial} presented a comprehensive adversarial sample comparison and discussed existing countermeasures to defend ASR. Sch\"{o}nherr \textit{et al.}~\cite{schoenherr2019} attacked ASR systems using inconspicuous adversarial perturbations in psychoacoustic hiding to lower acoustic signals under the threshold of human perception. Hence, imperceptible thus dangerous attacks can be implemented by altering audio signals. In the future, ASR has the potential to be broadly used to control Artificial Intelligence of Things (AIoT) devices~\cite{younis2021challenges,ma2019privacy,zhang2020empowering}, so ASR systems need to withstand security attacks. This paper focuses on inaudible voice command attacks and develops a multimodal defense method in ADAS.

\subsection{Audio Defense}
In the aforementioned audio attacks, researchers proposed related defense strategies. Carlini \textit{et al.}~\cite{carlini2016hidden} adopted simple defense methods, such as a beep (``the alert"), vibration (``the buzz"), and flashing LED indicators (``the light show"), to inform  of an attack. Diao \textit{et al.}~\cite{diao2014your} found that the microphones supporting multi-processing stopped the audio adversarial attacks automatically. 
% The same works on customised permission settings on a speaker's status checking. 
However, these proposed defense solutions led to low customer satisfaction. Sugawara \textit{et al.}~\cite{sugawara2020light} presented both software and hardware defense solutions. In terms of software, they advised adding an additional layer for authentication or utilizing voice control system device locations to prevent eavesdropping. Meanwhile, they suggested manufacturers to employ a multi-microphone array for sensor fusion because a laser cannot attack the microphone array simultaneously. In terms of hardware, they displayed a microphone design that used a silicon plate or a movable shutter to decrease the quantity of light reaching the diaphragm of the microphone~\cite{sugawara2020light, wang2015era}.

Zhou~\textit{et al.}~\cite{zhou2019defense} proposed a defense method against hidden voice commands on autonomous driving. That identified a live user speaking while breathing close to the microphone, and it used Gammatone Frequency Cepstral Coefﬁcients (GFCC) features to detect the location if voice commands came from the user. Moreover, He~\textit{et al.}~\cite{he2019canceling} introduced a signal transmitter to actively generate a special spectrogram using the passband, canceling inaudible-voice commands. Zhang~\textit{et al.}~\cite{zhang2017dolphinattack} used the Support Vector Machine (SVM) to classify between the genuine voice and inaudible voice by different spectrograms. However, existing defense methods, including signal processing, are not practical for identifying inaudible attacks in real-time, and it is difficult to collect inaudible command samples for machine learning training.

\subsection{Sensor Fusion and Trustworthiness}
Sensor fusion in ADAS is used to collect surrounding environments for an accurate full view of road conditions by multi-modal or multi-view data, which helps integrate a range of supplementary information to achieve a reliable driving experience with fewer false alarms. Moreover, heterogeneous sensor fusion for effective trajectory planning maintains an accurate state and estimates consistent positions. Sensor fusion in deep learning has become a prominent research topic multi-modal and has been adopted in relevant applications~\cite{summaira2021recent}. Fayyad \textit{et al.}~\cite{fayyad2020deep} surveyed deep learning sensor fusion on perception and its trends. Wang \textit{et al.}~\cite{wang2019survey} compared the advantages and disadvantages of different sensors and summarized fusion strategies. 
Joze \textit{et al.}~\cite{joze2020mmtm} introduced a Multimodal Transfer Module (MMTM) to fuse knowledge from multiple modalities in convolutional neural networks (CNN). MMTM can insert to different network architectures with minimum changes in gesture recognition, audio-visual speech enhancement, and action recognition. 
Compared to simple feature concatenation. Sterpu \textit{et al.}~\cite{sterpu2018attention} introduced the attention mechanism as the fusion strategy to automatically learn audio-vision alignment for enhanced representation in speech recognition tasks. 
The improvements under high-quality images were up to 30\% over acoustic modality alone in clean and noisy conditions. Xiong \textit{et al.}~\cite{xiong2021attackav} investigated multi-source adversarial sample attacks on autonomous vehicles. Specifically, Shen \textit{et al.}~\cite{shen2020GPSspoofing} analyzed off-road and wrong-way attacks by multi-sensor fusion under GPS spoofing. To the best of our knowledge, no sensor fusion has been used in-depth for the in-vehicle anti-inaudible voice command system.

Trustworthiness is vital for human-centered AI-based systems, allowing how machine learning impact users to trust model behaviours~\cite{toreini2020relationship}. Perceptions of trustworthiness reinforce AI system safety to be resilient and secure. In addition, trustworthy sensor fusion studies are in an early stage with a limited literature, which is challenging for multi-modal fusion in deep learning-based AI systems. Deep learning typically emphasizes probabilities even in achieving human-level performance, but individuals concern model reliability associated with the service such as an autonomous driving field. The deep learning model is an essential component of modern ADAS, and developers can trust the capabilities of the justifiable model in ADAS. On the other hand, reliability is of great significance to ADAS, and lack of interpretability results in trust problems in AI-based systems. And deterministic models are for certain predictions, which are not a comprehensive measurement for reliable trustworthiness~\cite{varsheny2019trustworthy}. Thus, assessing trustworthiness in deep learning becomes vital and its growing interest in interpretability. Model uncertainty as one aspect of trustworthiness inspires us to discover it in autonomous driving. Since providing one aspect of model uncertainty can be biased, this study explores Calibration Error~\cite{ovadia2019uncertainty} as an uncertainty metric and Monte-Carlo Dropout \cite{gal2016dropout} as uncertainty estimates to avoid biased model predictions. More details are presented in Section \ref{sec:Methodology}.

\section{Defense Framework}
\label{sec:Methodology}

\begin{figure*}[htbp]
\centering
\includegraphics[width=.9\textwidth]{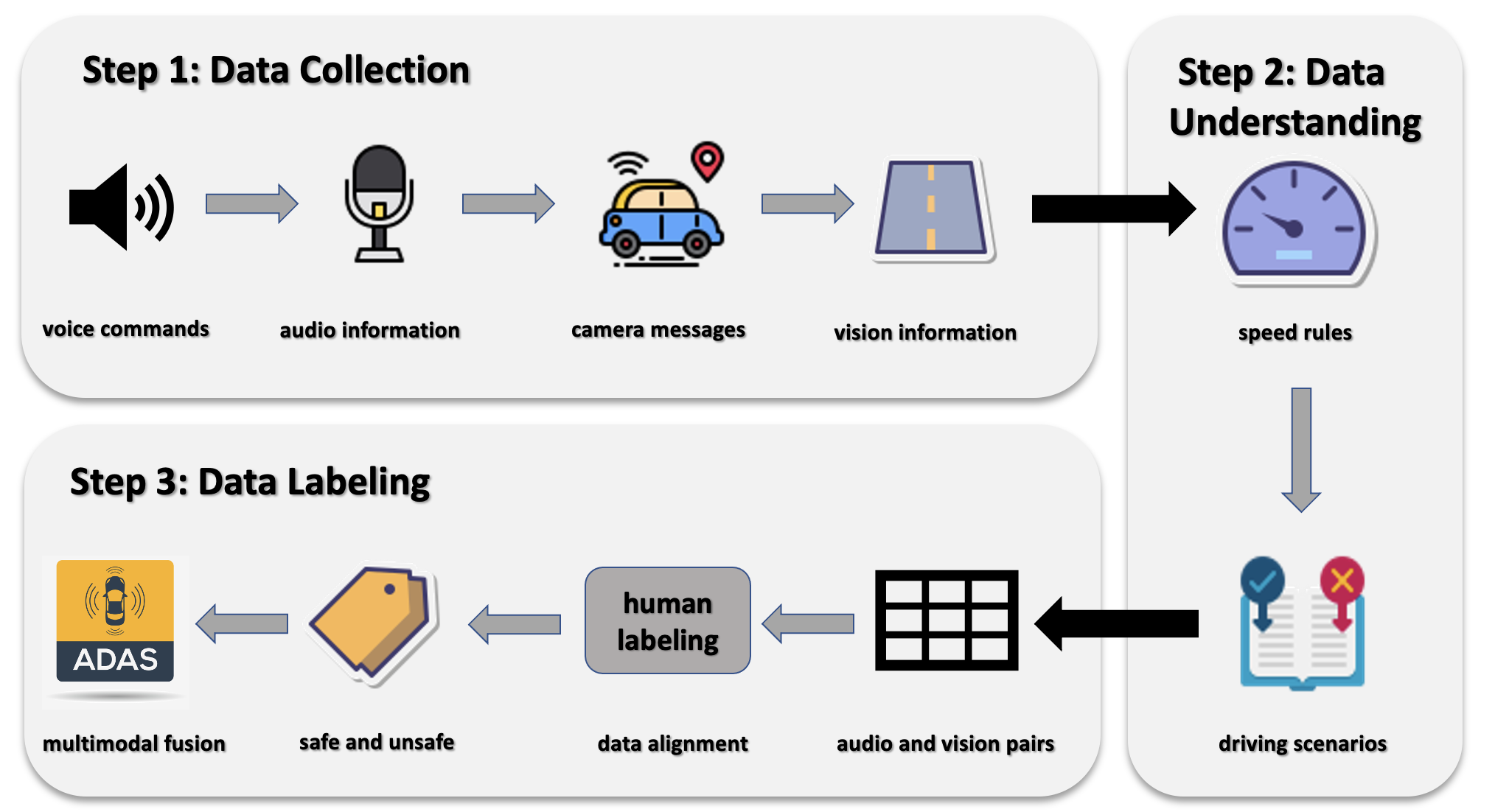}
\caption{The data labeling workflow}
\label{fig 2} 
\end{figure*}

Our study aims to develop a multi-modality fusion deep learning model integrating audio and vision. The model's responsibility uses visual information captured from multi-cameras to detect inaudible voice commands in ADAS. Audio inputs are semantically contrary to visual driving perceptions. Requirements at this phase focus on a binary classification: safe or unsafe vehicle maneuver. The proposed MFF data workflow is shown in Fig.~\ref{fig 2}, including data collection, data understanding and data labeling. 
Data collection collects various sources, such as gathering voice commands from a MEMS microphone and collecting driving views from front cameras. Data understanding is a systematic process of defining driving scenarios by speed. For example, the captured visual information consists of two driving scenarios: go and stop. If receiving an inaudible voice command ``stop", ADAS refuses immediately to execute this command to stop a vehicle when safely driving on the highway. In such case, the voice command conflicts with visual information in the driving scenario. In aggregation, we propose four rules to define audio and vision pairs:

\begin{itemize}
    \item Rule 1: While driving, the user says the ``stop" command, but ADAS using camera detection says no.
    \item Rule 2: While driving, the user says the ``stop" command, and ADAS using camera detection says yes.
    \item Rule 3: While stopping, the user says the ``go" command, but ADAS using camera detection says no.
    \item Rule 4: While stopping, the user says the ``go" command, and ADAS using camera detection says yes.
\end{itemize}

Data labeling aims to identify corresponding images for voice commands. 
We consider human knowledge as the oracle to align audio and vision information under driving scenarios. We introduce the data labeling in detail in Section \ref{sec:Experiment}.

From a safety perspective, ADAS makes correct real-time decisions as precise as a human driver within the required time limit. In these circumstances, the two driving challenge aims to be fulfilled in certain conditions: (1) how the proposed fusion model detects inaudible voice commands with audio-vision fusion; (2) how the fusion model in a making decision process is trustworthy to execute.

\textbf{Aim 1: Detect inaudible voice commands with audio and vision fusion attack.}
In Level-5 autonomous driving, perception modules detect driving environments and provide multimodal information to ADAS. The proposed ADAS method parses voice commands when receiving voice commands and executes motion planning. Meanwhile, an inaudible voice command attacks the driving vehicle, and its semantic content conflicts with the visual information from multi-cameras, confusing the ADAS. Once ADAS detects semantically conflicting visual information versus inaudible voice commands, it refuses to execute a command. 

\textbf{Aim 2: Perception of trustworthiness in the defense system}.
Multi-modal sensor fusion defense strategy can achieve a highly accurate predictive detection in ADAS. To assess the trustworthiness, model uncertainty indicates how reliable the prediction is and shows MFF defense system abilities and limitations for users. The uncertainty awareness reveals the defense system's decision boundaries.

\begin{table}[htbp]
\caption{Structure of VGGish}
\label{tab:VGGish}
\centering
\begin{tabular}{l|rrrr}
\hline\hline
    & \multicolumn{1}{l}{\textbf{Output Shape}} & \multicolumn{1}{l}{\textbf{Filters}} & \multicolumn{1}{l}{\textbf{Kernel}} & \multicolumn{1}{l}{\textbf{Activation}} \\ \hline
Conv2D+BN & 128, 44, 64                               & 64                                  & 3, 3                                & ReLU                                    \\
MaxPool2D & 64, 22, 64                                & -                                   & 2, 2                                & -                                       \\
MCDropout & 64, 22, 64                                & -                                   & -                                   & -                                       \\ \hline
Conv2D+BN & 64, 22, 128                               & 128                                 & 3, 3                                & ReLU                                    \\
MaxPool2D & 32, 11, 128                               & -                                   & 2, 2                                & -                                       \\
MCDropout & 32, 11, 128                               & -                                   & -                                   & -                                       \\ \hline
Conv2D    & 32, 11, 256                               & 256                                 & 3, 3                                & ReLU                                    \\
Conv2D+BN & 32, 11, 256                               & 256                                 & 3, 3                                & ReLU                                    \\
MaxPool2D & 16, 6, 256                                & -                                   & 2, 2                                & -                                       \\
MCDropout & 16, 6, 256                                & -                                   & -                                   & -                                       \\ \hline
Conv2D    & 16, 6, 512                                & 516                                 & 3, 3                                & ReLU                                    \\
Conv2D+BN & 16, 6, 512                                & 516                                 & 3, 3                                & ReLU                                    \\
MaxPool2D & 8, 3, 512                                 & -                                   & 2, 2                                & -                                       \\
MCDropout & 8, 3, 512                                 & -                                   & -                                   & -                                       \\ \hline\hline
\end{tabular}
\end{table}

\begin{table}[htbp]
\caption{Structure of Left and Right VGG16}
\label{tab:VGG16}
\centering
\begin{tabular}{l|rrrr}
\hline\hline
\textbf{} & \multicolumn{1}{l}{\textbf{Output Shape}} & \multicolumn{1}{l}{\textbf{Filters}} & \multicolumn{1}{l}{\textbf{Kernel}} & \multicolumn{1}{l}{\textbf{Activation}} \\ \hline
Conv2D    & 224, 224, 64                              & 64                                   & 3, 3                                & ReLU                                    \\
Conv2D    & 224, 224, 64                              & 64                                   & 3, 3                                & ReLU                                    \\
MaxPool2D & 112, 112, 64                              &  -                                   & 2, 2                                & -                                       \\
MCDropout & 112, 112, 64                              &  -                                   &  -                               & -                                       \\ \hline
Conv2D    & 112, 112, 128                             & 128                                  & 3, 3                                & ReLU                                    \\
Conv2D    & 112, 112, 128                             & 128                                  & 3, 3                                & ReLU                                    \\
MaxPool2D & 56, 56, 128                               &  -                                   & 2, 2                                & -                                       \\
MCDropout & 56, 56, 128                               & -                                 & -                                   & -                                       \\ \hline
Conv2D    & 56, 56, 256                               & 256                                  & 3, 3                                & ReLU                                    \\
Conv2D    & 56, 56, 256                               & 256                                  & 3, 3                                & ReLU                                    \\
Conv2D    & 56, 56, 256                               & 256                                  & 3, 3                                & ReLU                                    \\
MaxPool2D & 28, 28, 256                               &   -                                  & 2, 2                                & -                                       \\
MCDropout & 28, 28, 256                               &   -                                  &    -                                & -                                       \\ \hline
Conv2D    & 28, 28, 512                               & 512                                  & 3, 3                                & ReLU                                    \\
Conv2D    & 28, 28, 512                               & 512                                  & 3, 3                                & ReLU                                    \\
Conv2D    & 28, 28, 512                               & 512                                  & 3, 3                                & ReLU                                    \\
MaxPool2D & 14, 14, 512                               &  -                                   & 2, 2                                & -                                       \\
MCDropout & 14, 14, 512                               &  -                                   &  -                               & -                                       \\ \hline
Conv2D    & 14, 14, 512                               & 512                                  & 3, 3                                & ReLU                                    \\
Conv2D    & 14, 14, 512                               & 512                                  & 3, 3                                & ReLU                                    \\
Conv2D    & 14, 14, 512                               & 512                                  & 3, 3                                & ReLU                                    \\
MaxPool2D & 7, 7, 512                                 &  -                                   & 2, 2                                & -                                       \\
MCDropout & 7, 7, 512                                 &    -                                 &   -                                 & -                                       \\ \hline\hline
\end{tabular}

\end{table}

\begin{figure*}[htbp]
\centering
\includegraphics[width=.9\textwidth]{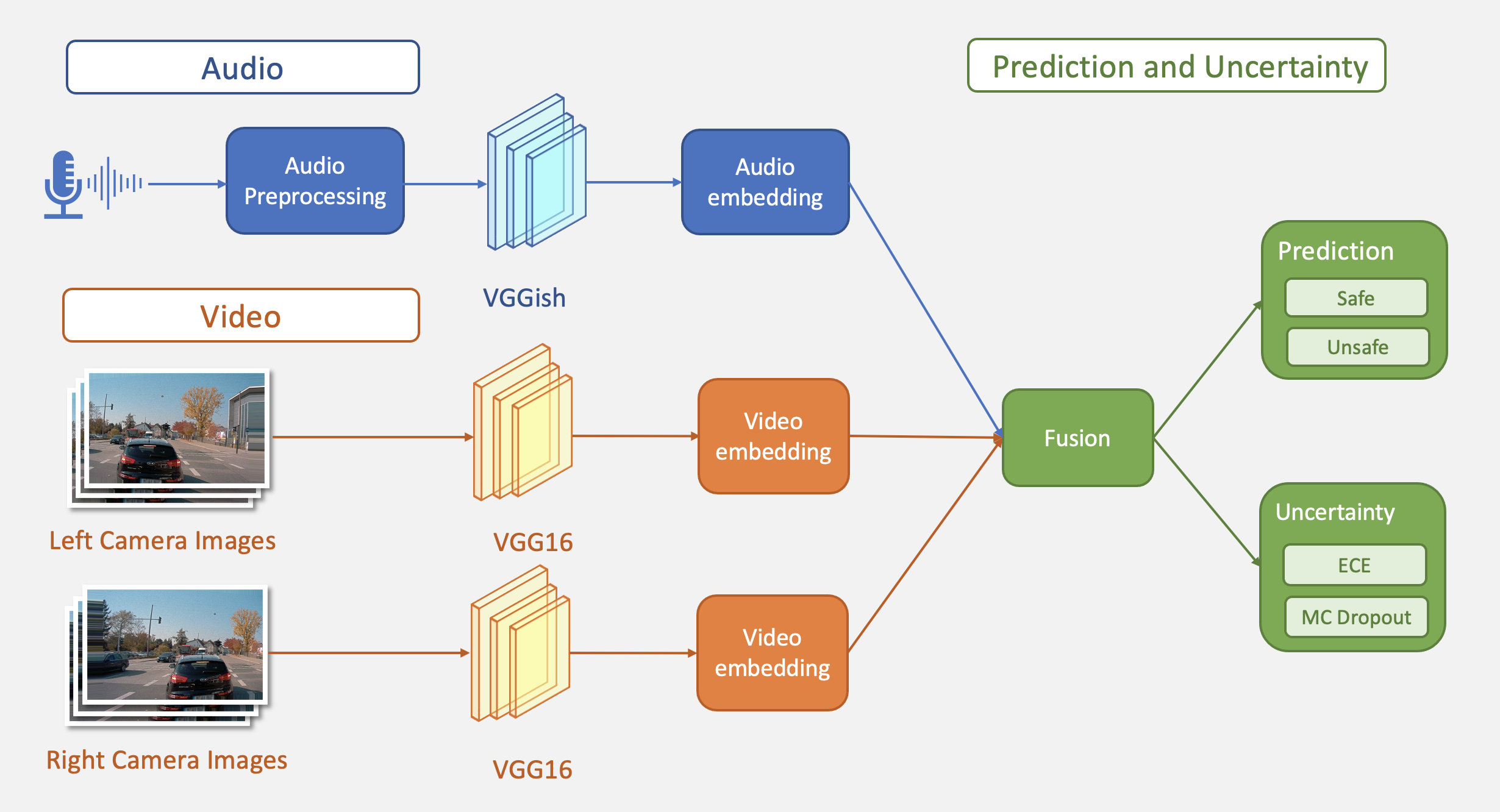}
\caption{The Parallel Fused Neural Network presents a visualization of the defense system architecture. It aims to integrate information from both modalities and build trustworthiness.}
\label{fig 3} 
\end{figure*}

\subsection{Overview of MFF}
\begin{figure*}[htbp]
\centering
\includegraphics[width=0.95\textwidth]{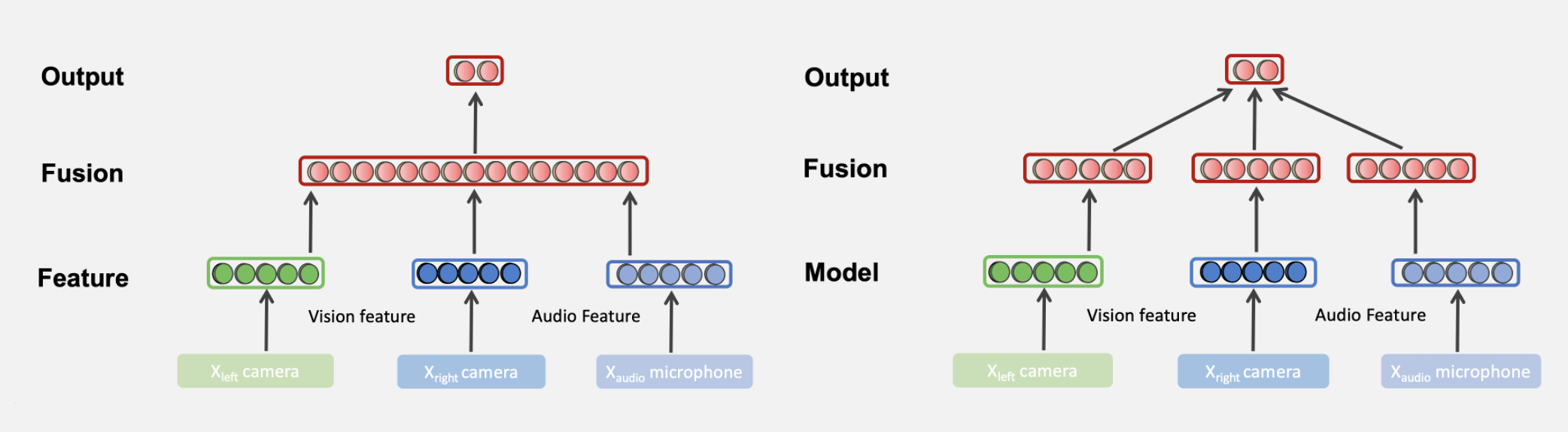}
\caption{The audio-vision early fusion and late fusion model architecture.}
\label{fig 4} 
\end{figure*}

An illustration of this multimodal fusion defense system architecture is shown in Fig.~\ref{fig 3}. It is explicitly designed to explicitly combine audio and video signals to be extracted as feature fusion. To introduce the proposed MFF method, we customize two CNN components that are from VGG16 \cite{simonyan2015vgg16} and VGGish \cite{aren2017vggish}. In the audio part, audio signals for the defense system are processed from waveforms into features, such as Mel-spectrogram \cite{shen2018spectrogram} and Mel-Frequency Cepstral Coefficients (MFCC) \cite{muda2010mfcc}. Then, the customized VGGish is used to transfer audio features into semantic embeddings. In the video part, image embeddings from the left and right cameras are combined by a customized VGG16 model. The architectures of customized VGGish and VGG16 are shown in Tables~\ref{tab:VGGish} and \ref{tab:VGG16}, respectively. Audio and video embedding are standardized before being analyzed by two modalities of each subnetwork. Each subnetwork adapts convolutional blocks with convolutional layers, Monte-Carlo Dropout layers and batch normalization layers if required. Depending on the fusion arrangements, the hybrid fusion stage is categorized into early fusion or late fusion~\cite{d2015review} as indicated in Fig.~\ref{fig 4}. The early fusion strategy assumes that multiple modalities to be processed independently have correlated features and involve concatenation. The late fusion is to independently process separate modalities and ensemble scores of model classifiers in the decision-making stage. To defend against inaudible voice command attacks, we compare the two multimodal fusions with proposed driving scenarios on the synthetic dataset.

\subsection{Fusion Strategy and Model Uncertainty}
In this phrase, multimodal early fusion is to fuse semantic embedding between audio and vision. This fusion typically employs supervised deep learning for contextualised representations to find the best mapping function $f_{\theta}:(x_{\text{audio}}, \, x_{\text{left}}, \, x_{\text{right}})  \rightarrow $  \ $y$,  %$label$ among dataset $\mathcal{S}=\left\{\left(x_{1}, y_{1}\right), \ldots, \left(x_{n}, y_{n}\right)\right\}$, 
where each sub network updates model parameters $\theta$ through $\mathcal{L}_{\mathrm{C}}$ binary cross-entropy loss function. By definition,  $y$ is adaptive for feature-fusion to concatenate $x_{\text{audio}}$ from audible inputs with $x_{\text{left}}$ and $x_{\text{right}}$ camera images from visual inputs. Then minimization process $\theta^{\star}$ subjected to network parameters is as follows.

\begin{equation}
\theta^{\star} = \underset{\theta}{\arg\min}[\mathcal{L}_{\mathrm{C}} \left(f_{\theta}(x_{audio} \mathbin\Vert x_{left} \mathbin\Vert x_{right}), y\right)]
\end{equation}

On the other hand, multimodal late fusion utilizes multiple model classifiers to derive the detection results. 
The decision fusion technique incorporates individual networks with the binary classification and cross-entropy loss function to train the model parameters $\alpha$ and $\beta$ in the corresponding models. 
The fusion scheme combines $f_{\alpha}(x\_{audio}, y)$ for audio detection and $f_{\beta}(x\_{left} \mathbin\Vert x\_{right}, y)$ for multi-view vision detection. 
The optimization of $\alpha^{\star}$ and $\beta^{\star}$ is outlined as follows.

\begin{equation}
\alpha^{\star}, \beta^{\star} = \underset{\alpha, \beta}{\arg\min}[\mathcal{L}_{\mathrm{C}} \left(f_{\alpha}(x_{\text{audio}}, y), f_{\beta}(x_{\text{left}} \mathbin\Vert x_{\text{right}}, y\right))] 
\end{equation}

\textbf{Expected Calibration Error (ECE)}.
For model uncertainty, intuitive metrics are needed. Expected Calibration Error (ECE) \cite{nixon2019measuring,zhang2020mix,kumar2019verified} differentiates between average actual accuracy and predicted confidence, which is a scalar summary of confidence calibration for model uncertainty quantitatively. ECE measures the average of calibration errors between the actual accuracy and confidence (predicated probabilities) across intervals into bins. The definition is summarized as follows:

\begin{equation}
\text{ECE} = \sum_{m=1}^{M}\frac{\left | B_m \right |}{n}\left | \text{acc}(B_m)-\text{conf}(B_m) \right |
\end{equation}

\noindent where $n$ is the total number of samples, \textit{M} is the number of bins which evenly split the probability interval [0,1] and $| B_m |$ is the amount of samples in each non-empty bin. $acc(B_m)$ is the ratio of correctly predicted items in the bin and $conf(B_m)$ is the average prediction in the bin $B_m$. In relation to the degree of uncertainty, trust calibration considers how a fused deep learning model is underconfident or overconfident. In other words, the lower the ECE, the better the trusted model. Miscalibration beyond the scope of model capabilities leads to distrust or overtrust. These improper trust behaviors can cause disastrous accidents. Hence, ECE is suitable to show inappropriate trust in ADAS decision-making.

\textbf{Monte-Carlo Dropout.}
Dropout is a technique for addressing overfitting problems in deep learning. This method randomly disables neuron units from the neural network when training and it is used layer by layer and applied to any hidden layers. It prevents neurons from co-adapting too much to increase model accuracy \cite{srivastava2014dropout}. Additionally, Monte-Carlo Dropout~\cite{gal2016dropout,liu2019universal,nair2020exploring} are based on Bayesian approximation in deep Gaussian processes. It refers to a stochastic forward-passing network to calculate the mean values of the predicted results in iterations. Monte-Carlo Dropout is an effective method to create a distribution of multiple iteration predictions for each test data. Multiple iterations with Monte-Carlo dropout are analogous to acquiring predictions from an ensemble network, as Bayesian posterior inference is difficult to derive the exact answer but can approximate it. This observation allows us to do an extensive study that evaluates the posterior distribution to assess the uncertainty estimation for trustworthiness. Experiments require a customized Monte-Carlo Dropout class inherited from the regular Dropout and then integrated into the deep neural network architecture.

\section{Dataset and Experiments}
\label{sec:Experiment}
\subsection{Dataset} 

Since there are no multi-modal datasets containing driving scenarios and proper audio commands, we create the development dataset consisting of a set of driving videos from an Audi A2D2~\cite{geyer2020a2d2} and audio from the Google Speech Command dataset~\cite{warden2018speechcommands}. A2D2 is an autonomous driving dataset that covers more driving scenarios and captures large static objects for object detection. It also has 3D bounding box annotations and instance segmentation annotations for all available traffic participants, such as vehicles and pedestrians. The Google Speech Command dataset consists of single spoken words for keyword spotting systems, and audios are recorded in approximately one-second recordings instead of a long speech. We select the speech commands ``go" and ``stop" from this dataset for our experiment.

\textbf{Audio Preprocessing}. Audio signals are recorded as waveforms, while the machine cannot directly process raw waveforms. Thus, feature extraction is a necessary step to convert them into understandable formats. The original VGGish accepts audio features by Mel-spectrogram. Mel-spectrogram is designed to convert signals from a time domain into the frequency domain, including sliding window and frame segment, Discrete Fourier Transform (DCT). Another popular waveform processing is to use MFCC. MFCC usually contains a sliding window of 25 ms, which separates the signals into short frames. Then it uses DCT, Mel-filter bank, Log and Discrete Cosine Transform. In this experiment, we use the Mel-spectrogram and MFCC as audio inputs to compare. The Python Librosa package extracts features from the voice command waveforms of ``go" and ``stop". The audio feature extraction for Mel-spectrogram uses a sampling rate with 16Kbit/s and DCT type-2 normalization. Feature extraction for MFCC mainly uses the same parameters setting while the number of filters in MFCC is 24 with a 25 ms sliding window and 10 ms frameshift.

\begin{table}[htbp]
\caption{Driving Scenarios}
\label{tal:3}
\centering
\scalebox{1.1}{
\begin{tabular}{c|ccc}
\hline\hline
\textbf{Scenarios} & \textbf{\begin{tabular}[c]{@{}c@{}}Audio\\ command\end{tabular}} & \textbf{\begin{tabular}[c]{@{}c@{}}Video \\ decision\end{tabular}} & \textbf{\begin{tabular}[c]{@{}c@{}}ADAS\\ decision\end{tabular}} \\ \hline
Scenario 1         & Go                                                             & Cannot go (0 km/h)                                                            & Unsafe                                                           \\
Scenario 2         & Go                                                             & Can go (5-10 km/h)                                                            & Safe                                                             \\
Scenario 3         & Stop                                                           & Can stop (15-20 km/h)                                                         & Safe                                                             \\
Scenario 4         & Stop                                                           & Cannot stop (25-$\infty$ km/h)                                                        & Unsafe                                                           \\ \hline\hline
\end{tabular}}
\end{table}

\begin{figure}[htbp]
\centering
\includegraphics[width=.49\textwidth]{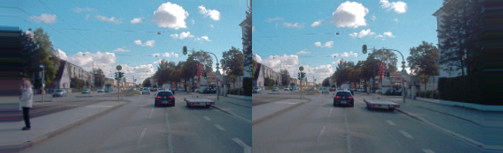}
\caption{Example of left and right camera visual information}
\label{fig 5} 
\end{figure}

%  Each image is down-sampled from 1920*1208 to 288*181.
\textbf{Video Preprocessing}. 
We provide the visual information pairs from the A2D2 dataset, considering the ``go" and ``stop" driving video clips.
As the A2D2 dataset contains speed labels, we select the four speed range to capture images from the A2D2 video, as shown in Table~\ref{tal:3}. These images are classified into four driving scenarios according to the velocity with the two voice commands. Images captured by both left and right cameras are aligned from video inputs, as shown in Fig~\ref{fig 5}.
If the voice command conflicts with the scenario in the driving image, the defense system identifies it as unsafe and rejects the command, and vice versa. We introduce Human-in-the-loop (HITL)~\cite{nunes2018hitl}, which enhances human labeling for MFF to select the right driving image and voice command pairs to train the fusion deep learning model. Human labeling provides knowledge and experience for the development dataset. Eventually, the development dataset has the left and right road images in a scene collected by two front cameras, corresponding to the voice command. We use an audio-aligned image to group these pairs, including safe and unsafe situations, and the synthetic training dataset is composed of 4000 pairs.

\subsection{Experiment Setup}
All subnetworks in MFF use ReLU activation and cross-entropy loss function. We adopt the fine-tuning strategy by a pre-training model from ImageNet \cite{kornblith2019better} for a sophisticated multi-modality model. Adam \cite{kingma2015adam} is chosen as the optimizer for training, with a batch size of 64 and a learning rate of 0.01. The dropout rate in each dropout layer is 0.5. Experiments perform 10-fold cross-validation, and we set 100 epochs to save the best-performing model. Early stopping is applied to avoid over-fitting with ten patience. All experiments are implemented in Keras with the TensorFlow backend. These experiments are conducted with a Windows operation system, Intel i7-9700 3.0 GHz, 64 GB of memory, and two NVIDIA RTX 2080Ti GPUs, which take up to 10 hours of training. To evaluate the feasibility of MFF against inaudible voice command attacks, evaluation metrics include accuracy, precision, recall, and F1-measure. Model uncertainty is evaluated by ECE.

\section{Performance Evaluations}
\label{sec:Evaluation}

The fusion model performance is evaluated to distinguish safe or unsafe decisions during an inaudible voice command attack. We conduct the ablation study that is momentous to understand early and late fusion methods influenced by different feature extractions and expand on trustworthy findings. To assist our analysis, we show precision, recall, and F1-score by Mel-spectrogram and MFCC inputs in Table~\ref{tal 3}. The early fusion detection accuracy on the overall test data is 92.25\% by Mel-spectrogram and 91.50\% by MFCC. The late fusion detection accuracy is 90.00\% by Mel-spectrogram and 92.00\% by MFCC. Hence, our results show that the early fusion method outperforms the late fusion in accuracy for Mel-spectrogram, but lower accuracy using MFCC. Under the same audio feature extraction, this result also demonstrates that improving visual features is crucial to matter audio and vision joint representation tasks. We observe that these traffic sign images are limited and have fewer variations, as a simple traffic sign cannot retrieve greater diversity of camera inputs from real-world driving. Two cameras continuously allow for richer visual information as one camera with traffic signs is too limited to learn a reliable visual semantic embedding. Subsequently, one traffic sign from one camera matching a specific inaudible voice command is unreliable; if that one camera is under adversarial attacks, the fusion result is not trusted. Therefore, a single camera is unreliable for inaudible attacks and does not offer a practical security solution.

\begin{table}[htbp]
\footnotesize
\caption{Evaluation and Expected Calibration Error of MMF Model}
\label{tal 3} 
\scalebox{1.5}{
\centering
\resizebox{60mm}{6mm}{
\begin{tabular}{c|cccccc}
\toprule

\textbf{Fusion Method} &  \textbf{Features} & \textbf{Accuracy} & \textbf{Precision} & \textbf{Recall} & \textbf{F1} & \textbf{ECE} \\ 

\cmidrule(lr){1-7}\ %\cmidrule(lr){3-5}\cmidrule(lr){5-7}
\multirow{2}*{Early Fusion} & Mel-spectrogram   & \textbf{92.25\%}    & 89.00\%  & 95.00\%  & 92.00\%   & \textbf{6.21}    \\ 
                            &  MFCC             & 91.50\%  & 91.00\%  & 91.00\%   & 91.00\%  & 5.71    \\ \hline\
\multirow{2}*{Late Fusion} & Mel-spectrogram    & 90.00\%  & 90.00\%  & 90.00\%  &  90.00\%  & \textbf{6.97}  \\ 
                           &  MFCC              & 92.00\%  & 92.50\%  & 92.50\%  &  92.50\%  &  3.01 \\ 

\bottomrule
\end{tabular}
}}
\end{table}

\begin{figure}[htbp]
\centering
\includegraphics[width=.5\textwidth]{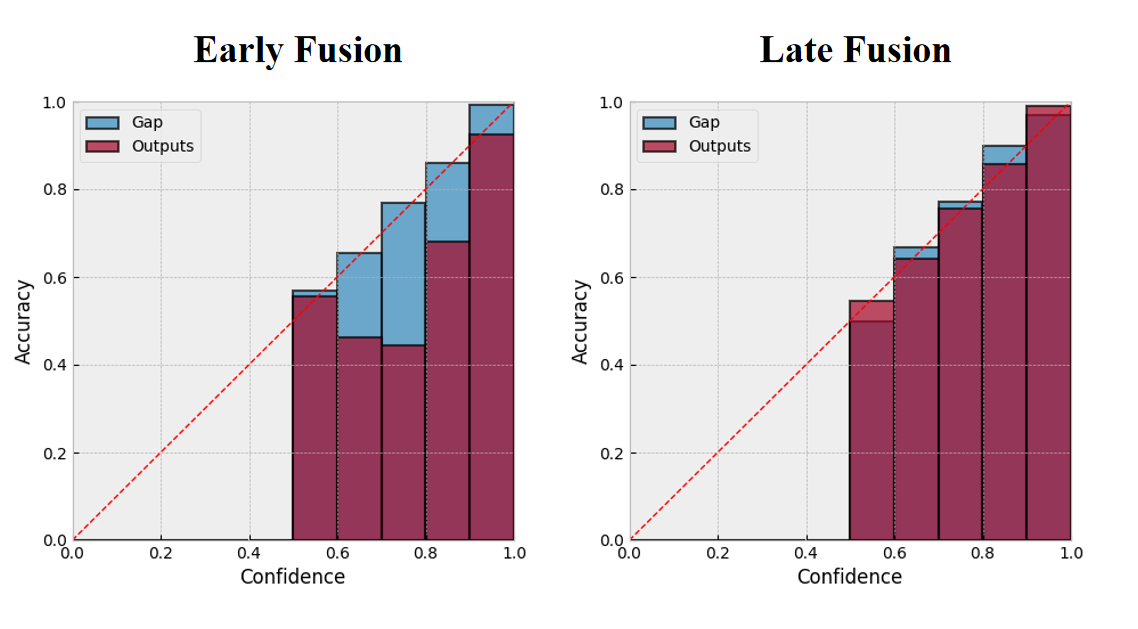}
\caption[Calibration\_Error]{The reliability diagram illustrates the distribution of ECE. The red bar is the prediction and blue bar is the gap. Early fusion is left and ate fusion is right.}
\label{fig 6} 
\end{figure}

Besides default evaluation metrics, we use selected fused models to evaluate the uncertainty. We use ECE values for reliability measurement purpose. The ECE column in Table~\ref{tal 3} shows Mel-spectrogram and MFCC for the early fusion model as 6.21 and 5.71, respectively. Although Mel-spectrogram benefits the accuracy, the detection reliability is overestimated by ECE compared to MFCC. Regarding the late fusion, the ECE values are 6.97 by Mel-spectrogram and 3.01 by MFCC. The late fusion by MFCC achieves not only a higher accuracy but also lower ECE. We experimentally train late fusion in reducing model uncertainty by vastly fewer model parameters and achieve similar accuracy as that of early fusion. Hence late fusion benefits audio-vision multimodal fusion tasks in low-resource IoT devices (e.g., autonomous vechiles). Calibrated errors are vital for model interpretability to establish trustworthiness~\cite{guo2017calibration}. To further investigate classification decisions, we conduct extensive analysis on calibration errors and we use reliability diagram for a visual representation to model calibration. The reliability diagram depicts plotting bars to indicate the calibrated errors in each bin and the gap shows the error between average confidence and accuracy. In Fig.~\ref{fig 6}, the more blue gaps are, the more ECE errors. Meanwhile the average confidence in early fusion is greater than its accuracy, the average confidence of late fusion closely matches its accuracy.

{Fig.~\ref{fig 7} is the confidence histogram with 10 bins to show the distribution of prediction confidence, including early fusion (top) and late fusion (bottom).
The confidence histogram in the test data presents that the majority of predictions by fusion models had a confidence level of greater than 0.9. The black vertical line shows the overall true accuracy, and the black dotted vertical line presents average prediction confidence. The average confidence of the early fusion is considerably higher than its accuracy, which is considered an untrustworthy model. While the average confidence of late fusion closely matches its expected accuracy, which is the well-calibrated model. It indicates that MFF with late fusion successfully identifies inaudible voice command attacks with strong confidence.

\begin{figure}[t]
\centering
\includegraphics[width=.47\textwidth]{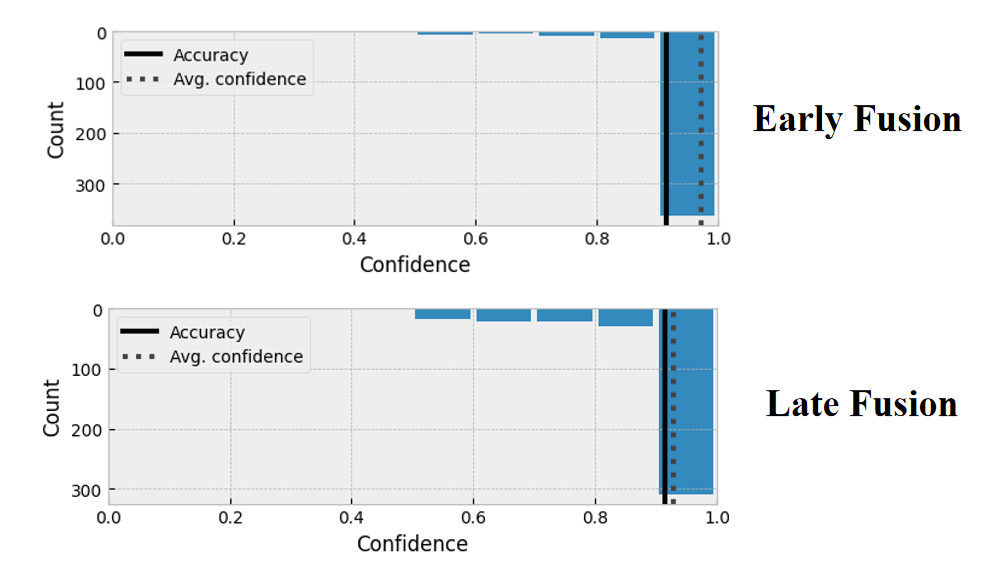}
\caption{The confidence histogram shows accuracy and average confidence in prediction. Count indicates the number of audio-image pairs in each bin, and confidence is the output.}
\label{fig 7} 
\end{figure}

Furthermore, the Monte Carlo Dropout provides the prediction uncertainty by the parameter distribution. 
With different parameter distributions, the model can output multiple predictions for each instance at inference time. 
Fig.~\ref{fig 8} shows the prediction accuracy distribution by Monte-Carlo Dropout in the test set and its ensemble accuracy using iteration of 100, where the early fusion occurs on the left-hand side and the late fusion on the right-hand side. 
For the early fusion model, the average Monte-Carlo accuracy is 91.55\%, and the red line is the Monte-Carlo ensemble by one test sample with an accuracy of 92.25\%. For the late fusion model, the average Monte-Carlo accuracy is 91.49\%, and the Monte-Carlo ensemble is 93.75\%. 
From the prediction uncertainty estimation, the above fusion models by the Monte-Carlo Dropout method obtained an uncertainty interval without adjusting the existing model. Thus, MFF aggregates prediction estimation for the uncertainty without extra effort. 
However, due to its high computational complexity, the Monte-Carlo Dropout method is only suitable for testing in the development phase.

\begin{figure}[t]
\centering
\includegraphics[width=.51\textwidth]{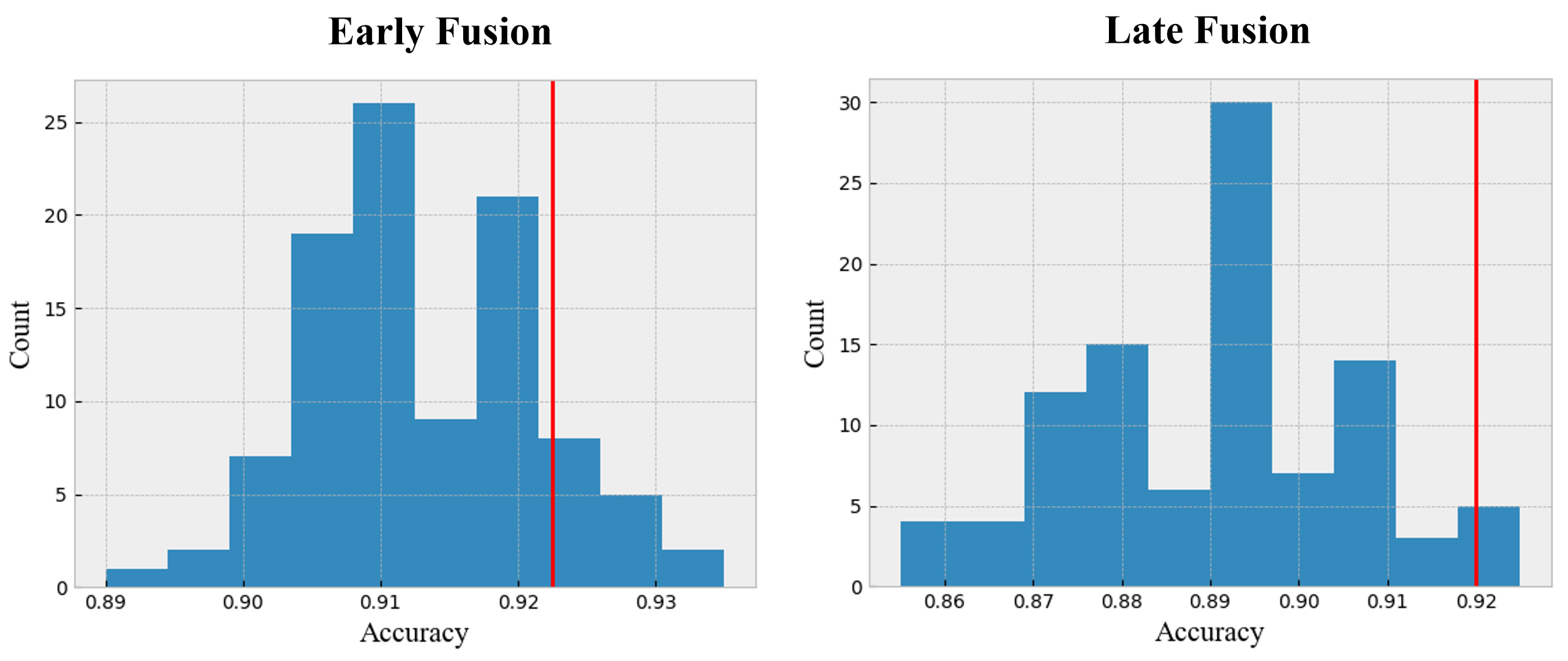}
\caption[Calibration\_Error]{The Monte-Carlo Dropout presents a scalable approximated predictive distribution. The red line shows the Monte-Carlo ensemble accuracy during test. The count indicates the number of iterations for the corresponding accuracy range.}
\label{fig 8} 
\end{figure}

\section{Discussion}
\label{sec:Discussion}

The proposed multimodal defense strategy is conducted to be effective in detecting inaudible voice command attacks and explores the audio-vision fusion model uncertainty. The estimating decision boundary for the fusion model provides clarity to stakeholders and underpins the critical decisions by ADAS. This is also highly desirable to represent the trustworthiness of AI-based ADAS in real-world IoT applications. Notably, compared to traffic signal model, it assumes that a single camera is not available to understand the rich visual information sufficiently. An effective and practical approach is to add more cameras for essential context-awareness, which is semantically consistent between audio \& vision modalities. We can extend to the use of multiple camera for detecting inaudible voice attacks, but this conflicts with the multi-view decision, which is not the primary purpose of our study. In addition, the reliability of this dataset is impacted by limited multi-view data, and they need to be from the same data distribution to train the fused model. However, such training data differs from sensor inputs in the underlying real-time data distribution, decreasing accuracy. Another problem is relying on humans to label audio-vision data, which might restrict encoding human knowledge to make ADAS trustworthy.

From model uncertainty aspect in the trustworthiness, substantial experiments help improve robustness and interpretability in multimodal models as model uncertainty is a way to adjust the training stage and generate reliable predictions. Recent work~\cite{nixon2019measuring} proposed calibration errors to recalibrate models successfully but failed in the evaluation stage. The limitation by results is constrained to the balance between predictions where there is even better performance on inference and true confidence. One potential solution is to set a threshold for calibration errors with desired confidence and then move forward to predictions that ensure the decisions are calibrated. The deep learning model usually relies on overestimated prediction probabilities if not well-calibrated. This setup is crucial to the sensitive multi-modal deep learning application. It should be noted that model calibration for model uncertainty does not end trustworthy machine learning problems. 

The proposed MFF is a practical method for decisive protection in reducing inaudible voice command attacks associated with visual responsibility. The advantage of the MFF strategy is its high accessibility because MFF does not require inaudible voice command samples that are difficult to collect as training data. One disadvantage is that the proposed method does not consider factors like GPS trajectory and object distance. Another shortcoming is that low-quality visual information by cameras leads to motion blur due to different weather conditions.

\section{Conclusion and Future Work}
\label{sec:Conclusion}

This paper proposes MFF as an empirical defense framework to detect inaudible voice command attacks in autonomous driving. 
MFF uses joint audio and vision embedding to produce cross-modal representations in a supervised way to correct driving decisions. 
In addition, extensive experiments fill a large gap for the multimodal fusion on trustworthy problems, especially by model uncertainty. Our results demonstrate that deep learning models need to trade off between accuracy and robustness.
The explicit model uncertainty analysis leads to understanding of joint audio and vision tasks. 
It also shows the reliability robustness of the fusion model and significantly empowers context awareness of voice-activated IoT security, which is deployed for safety-critical applications. 
Besides, using audio and video modalities with an early concatenation restricts its fusion capacity. 
Future work will be directed toward new frontiers in audio and vision contrastive pre-training.
Furthermore, future work needs to compare the evolutionary multimodal transformers in audio and vision tasks and explore the reliability of essential features toward designing a robust multimodal defense model for autonomous driving.

\appendices

\section*{Acknowledgement}
We appreciate anonymous reviewers for their constructive feedback and valuable discussions. This work is in part supported by the CSIRO Data61 Collaborative Research Project (CRP) C020996, CSIRO Data61 Topup 194981059 and Australian Research Council Linkage Project (ARC) LP190100676. This work is also supported by the Taishan Scholars Program under Grant TSQN 202211214 and Shandong Excellent Young Scientists Fund Program (Overseas) No.~2023HWYQ-113.

\ifCLASSOPTIONcaptionsoff
  \newpage
\fi

\bibliographystyle{IEEEtran}
\bibliography{IEEEabrv,references}

\end{document}